%% file: ska_interacting_binaries.tex
\documentclass[a4paper,11pt]{article}
\usepackage{aaskaiid}
\usepackage{orcidlink}
\newcommand{\AAstar}{AA$^{\ast}$}
\newcommand{\perbeam}{\,beam$^{-1}$}

\title{Accreting Compact Object Binaries with the SKA}
\ShortTitle{Accreting Compact Object Binaries with the SKA}

\author[1]{Aru Beri\orcidlink{0000-0003-3753-3102}}
\ShortName{A. Beri et al.}
\author[2]{Francesco Carotenuto\orcidlink{0000-0002-0426-3276}}
\author[3,4\dagger]{Rob P. Fender\orcidlink{0000-0002-5654-2744}}
\author[5,\dagger]{James C. A. Miller-Jones\orcidlink{0000-0003-3124-2814}}
\author[6]{Sara Motta\orcidlink{0000-0002-6154-5843}}
\author[7]{Valeriu Tudose\orcidlink{0000-0001-5317-220X}}
\author[8]{Jakob van den Eijnden\orcidlink{0000-0002-5686-0611}}
\author[]{the Transient Science Working Group}

\affiliation[1]{Indian Institute of Astrophysics, Koramangala II Block, Bangalore-560034, India}
\emailAdd{aru.beri@iiap.res.in}
\affiliation[2]{INAF-Osservatorio Astronomico di Roma, Via Frascati 33, I-00078, Monte Porzio Catone (RM), Italy}
\emailAdd{francesco.carotenuto@inaf.it}
\affiliation[3]{Department of Physics, University of Oxford, Denys Wilkinson Building, Keble Road, Oxford OX1 3RH, UK}
\emailAdd{rob.fender@physics.ox.ac.uk}
\affiliation[4]{Department of Astronomy, University of Cape Town, Private Bag X3, 7701 Rondebosch, South Africa}
\affiliation[5]{International Centre for Radio Astronomy Research -- Curtin University, GPO Box U1987, Perth, WA 6845, Australia}
\emailAdd{james.miller-jones@curtin.edu.au}
\affiliation[6]{Istituto Nazionale di Astrofisica, Osservatorio Astronomico di Brera, via E.\,Bianchi 46, 23807 Merate (LC), Italy}
\emailAdd{sara.motta@inaf.it}
\affiliation[7]{Institute of Space Science - INFLPR Subsidiary, Atomistilor 409, Magurele, Ilfov, RO-077125, Romania}
\emailAdd{tudose@spacescience.ro}
\affiliation[8]{Anton Pannekoek Institute, University of Amsterdam, Science Park 904, 1098 XH, Amsterdam, NL}
\emailAdd{a.j.vandeneijnden@uva.nl}
\affiliation[\dagger]{Chapter co-ordinator}

\abstract{Accreting binary systems provide a time-resolved view of the accretion and ejection processes that are seen in all classes of compact objects, from stellar-mass to supermassive. They allow us to study the launching of relativistic jets on human timescales, probing how the jets are coupled to the underlying accretion flow, and providing unique insights that can be extended to supermassive black holes via the well-established mass scale invariance. Comparative studies of jets from different classes of compact objects allow us to determine the impact of mass, spin, magnetic fields, and stellar surfaces on the launching of jets. The jets from black hole X-ray binaries provide probes of the Galactic black hole population, and an important source of feedback to the surrounding interstellar medium. While existing radio facilities (including the SKA precursors) have made great progress in this field in recent years, the SKA will enhance such studies via its high point source and surface brightness sensitivity, high angular resolution, and broad frequency coverage. This will enable the extension of our existing black hole studies to the fainter accreting neutron star and white dwarf population, the detection of previously-undiscovered systems in a low-luminosity quiescent state, and the extension of our studies to nearby galaxies, revealing rare, high-accretion rate systems that probe a key phase of black hole growth.}

\begin{document}
\include{journal-names}
\maketitle

\section{Introduction}
X-ray binaries (XRBs) consist of a stellar remnant (neutron star or black hole) accreting matter from a less-evolved donor star. They represent the end products of stellar evolution, providing us with an opportunity to conduct long-term electromagnetic studies of stellar remnants and stellar evolution, providing complementary information that can never be provided by the fleeting gravitational wave signals detected by the LIGO, Virgo and KAGRA interferometers.

While such stellar remnants are $10^3$--$10^4$ times closer than nearby AGN, they are also 8--9 orders of magnitude less massive, such that the achievable angular resolution with a given array is $10^4$--$10^6$ times coarser, in units of gravitational radii. Hence, we can never hope to resolve the event horizons or jet launching regions in XRBs, in the manner now routinely done for nearby AGN \citep{EHT2019a,Janssen2021}. However, XRBs provide a highly complementary probe of accretion physics and jet launching, opening up the time dimension to enable studies of the causal relation between inflow and outflow around accreting compact objects. Since black hole physics is expected to be scale invariant, the prevailing timescales should be proportional to the black hole mass, such that stellar-mass black holes in XRBs should vary on timescales millions to billions of times shorter than their supermassive analogues in AGN. The combination of spatially-resolved information from AGN and time-resolved information from XRBs will be needed to fully understand the physics of jet launching and kinetic feedback from black holes.

XRBs also provide ready-made laboratories for determining the effect on jet launching of mass, spin, a stellar surface, and a stellar magnetic field. By comparing jets from neutron star and even white dwarf systems with those from black holes, we can probe the impact of a stellar surface and magnetic field. The spins of neutron stars are often far easier to measure reliably than those of black holes \citep[e.g.][]{Choudhury2017}, and the full spectrum of accreting stellar-mass compact objects believed to launch jets spans a dynamic range of 40 in compact object mass \citep{Koerding2008,Miller-Jones2021}.

Furthermore, we can use XRBs as probes of the black hole population of the Milky Way. While a handful of black holes have been discovered via optical surveys over the past decade \citep{Giesers2018,Giesers2019,Lam2022,Sahu2022,ElBadry2023GaiaBH2,ElBadry2023GaiaBH1,GaiaBH3}, the majority of known Galactic black holes and black hole candidates reside in XRBs \citep{CorralSantana2016}. The radio jets of black hole XRBs can be used as astrometric probes to infer natal kicks and probe black hole formation mechanisms \citep[e.g.][]{Willems2005,Fragos2009}, and to reveal the presence of previously-undiscovered black holes \citep[e.g.][]{Strader2012,Chomiuk2013,Miller-Jones2015}.

As described in the remainder of this Chapter, the unrivalled sensitivity of the SKA will significantly enhance these studies, extending our previous knowledge of jets from black holes to their fainter analogues in neutron star and white dwarf systems. It will extend the range of studies of high accretion rate, radio-bright black holes out to nearby galaxies \citep[e.g.][]{Middleton2013}, and allow us to detect previously-unknown black holes in a faint quiescent state, where the bulk of the Galactic population resides \citep[e.g.][]{Gallo2006,Gallo2014}. Its excellent surface brightness sensitivity will allow us to search for the interactions between XRB jets and the surrounding interstellar medium, quantifying the kinetic feedback from a large enough sample of sources to enable population-level inferences. Finally, its broad frequency coverage will allow exquisite characterisation of the physical properties of jets, and how they evolve over time as the jets propagate through their environment.

\section{Jet launching}

The rapid time evolution of XRBs means that they can be used to determine the causal sequence of events leading to the launching of relativistically-moving jets. During the hard-to-soft X-ray state transition at the peak of an outburst, the accretion flow geometry changes, leading to significant changes in the X-ray flux and variability properties \citep{Belloni2010}, and the steady, compact jets seen during the rise phase of the outburst give rise to discrete, moving ejecta \citep{Fender2004}. 

The sensitivity of the SKA is not required to monitor these outbursts. However, by tracking the proper motions of those ejecta, initially on milliarcsecond scales with VLBI \citep[e.g][]{Miller-Jones2012}, and subsequently on larger scales with connected-element interferometers \citep[e.g.][]{Bright,Carotenuto_2021}, we can model their motions, and trace them back to the moment of zero separation, determining when the jets were ejected. With sufficiently dense X-ray coverage, we can determine the conditions in the accretion flow at the time of launch, and hence determine which changes in the accretion flow lead to the ejection of jets.

The limiting factor in such studies has to date been the accuracy with which we could determine the ejection dates, due to the difficulties of reliably imaging a variable, moving source with a sparse VLBI array, which violates the fundamental assumptions of aperture synthesis. The advent of new techniques \citep[e.g][]{Wood2025} has enabled the measurement of jet ejection times to within tens of minutes. The addition of SKA stations to existing VLBI arrays will improve the {\it uv}-coverage on the longest baselines, improving the performance of these techniques, and further augmenting the achievable time resolution. Coupled with its excellent low surface brightness sensitivity, it will allow us a longer lever arm with which to track the proper motions of ejecta.

To achieve these goals, the SKA would need to participate in few-hour VLBI observations within 1--2 days of the state transition occurring. Since sensitivity would not be an issue, this would make use of the subarraying capability of SKA, potentially using only a few long-baseline stations while core-only programs were conducting other science observations. Once the ejecta had expanded, occasional monitoring with the SKA-mid array over the course of several months would improve the time baseline available to fit for proper motions, as well as quantifying any deceleration in the jet motions. More detailed discussion of how SKA-VLBI can be used to study jet launching in X-ray binaries can be found in \citet{SKA_XRBs_VLBI}.

\section{Jet energetics and feedback}

\subsection{Discrete jets}

The powerful discrete jet ejecta launched during outbursts can be followed for months to years as they propagate out to parsec scales. Recent observations with MeerKAT have already revolutionised this field, revealing a plethora of discrete ejecta propagating to large distances from their central sources, with excellent coverage of the jet kinematics and radiation (e.g.\ \citealt{Bright, Carotenuto_2021, Bahramian_2023, Zhang_2025}). However, we still lack precise constraints on the physical parameters of these jets, such as their masses, Lorentz factors, and volumes. A key open problem is to quantify the jets’ total energy content (internal and kinetic). Measuring the jet energetics is essential not only for understanding the balance between accretion and ejection in BH XRBs, but also because of its implications for the jet composition, powering mechanisms, and the impact of feedback on the surrounding medium (e.g.\ \citealt{Fender_balance}).

The internal energy of the ejecta can be inferred from their synchrotron emission, provided that the size of the emitting region and the source distance are known, and assuming equipartition conditions \citep{Longair}. The size can be measured directly by spatially resolving the jet structure (e.g.\ \citealt{Rushton_2017}), inferred from the jet expansion speed and ejection timescale (e.g.\  \citealt{Russell_1535}), or estimated by associating peaks in the radio light curve with synchrotron self-absorption (e.g.\ \citealt{Fender_2019_equipartition}). Alternatively, multi-resolution radio interferometric observations can constrain the source size by measuring how much emission is resolved out on different baselines \citep{Bright}. On the other hand, the kinetic energy of the ejecta can be estimated through joint modelling of their kinematics and radiation, by measuring their angular separation from the core over time and applying physical models of jet–ISM interaction (e.g. \citealt{Wang_model, Hao, Steiner_xte}). Dense monitoring campaigns that capture how the jets decelerate in the ISM can provide important constraints on parameters such as the jet initial kinetic energy and Lorentz factor, and the density of the ISM through which it propagates \citep{Carotenuto_2024, Cooper_2025}.

In this context, the SKA will provide a fundamental advancement in our ability to measure and understand the energy budget of relativistic jets in X-ray binaries. Its sub-arcsecond angular resolution and high surface brightness sensitivity will make it possible to resolve multiple ejected components simultaneously and achieve a much denser coverage of jet kinematics, especially close to the moment of launch. Its improved \textit{uv}-plane coverage, high sensitivity, and broad multi-frequency capability will enable accurate measurements of jet morphology, size, and spectral evolution, which will be of prime importance to place robust constraints on the total jet energy. Moreover, the use of sub-arrays will allow simultaneous observations at multiple frequencies, providing broad spectral and temporal coverage of the evolving ejecta. For instance, the full \AAstar\ SKA-Mid array with 144 antennas operating in Band 2 will achieve a sensitivity of $\sim$3~$\mu$Jy beam$^{-1}$ over a 15-minute exposure, increasing the current MeerKAT sensitivity by a factor $\sim$7, enabling the detection of jet ejecta from many more BH XRBs in outburst, including observations of the faint, receding jet components, that are required to directly measure the jet speed and inclination angle. Overall, covering the full jet propagation with SKA-Mid will provide us with the first fully self-consistent estimates of the total jet energy and its evolution for this class of sources.

\subsection{Large-scale jet-induced structures}

As also observed in AGN, the jets in XRBs are able to inject energy and momentum into the surrounding interstellar medium (ISM). As these jets inflate over-pressured lobes and drive bow shocks, the resulting structures forming in the (shocked) ISM can be used as “calorimeters” of the time-averaged jet power \citep{Kaiser2004}. 

Until recently, only a few jet-inflated, parsec-scale structures had been identified around Galactic X-ray binaries—the archetype being the Cyg X-1 bow shock \citep{Gallo2005b}. MeerKAT’s surface-brightness sensitivity and arcsecond-resolution imaging have changed this picture, enabling multi-frequency observations that resolved faint (of order 0.1 mJy\perbeam) bow shock structures and lobes around multiple systems \citep[e.g.][]{Atri2025, Motta2025, Gasealahwe2025}, in some cases even revealing the structures' proper motion \citep{Mariani2025}. These data support calorimetric estimates of ambient density, age, and time-averaged jet power, showing that XRB jets can deposit energies comparable to—or exceeding—the radiative output over their lifetimes. As an SKA pathfinder, MeerKAT is offering only a glimpse of what the SKA will deliver, when such studies can be expanded into uniform, sensitive surveys of jet-driven feedback across the Galaxy.

The raw sensitivity and uv-coverage of the SKA will push down the surface brightness floor for diffuse emission so that fainter jet-inflated lobes and bow shocks around XRBs will become detectable in a few hours of integration. 
For a target at the equator, \AAstar\ will reach in 1 hr a Briggs-weighted sensitivity of 1.4\,$\mu$Jy\perbeam\ in Band 2, improving by an order of magnitude the sensitivity currently possible with MeerKAT. 

Additionally, the multi-frequency observing capabilities of the SKA, combined with its angular resolution, will spatially resolve the spectral index across the jet-induced structure, thus mapping the emission processes at play. 
The above configuration will provide an angular resolution of approximately $2.1\times1.8$ arcsec$^2$ in Band 2, and $0.08\times0.07$ arcsec$^2$ in Band 5b.

The SKA’s survey speed and imaging fidelity, combined with its large field of view, will enable critical tests that are currently sample-limited. Uniform mosaics will identify new bow shock structures and jet-inflated lobes independent of prior X-ray activity, reducing selection biases toward persistent or well-studied sources. Follow-up observations of jet structures on different time-scales (from days to years) will eventually connect the accretion history of a system to long-term energy deposition. 

The SKA will transform single-object “case studies” into a statistically robust census of jet energetics in the Milky Way. This will quantify how often, how strongly, and under which environmental conditions stellar-mass jets couple to the ISM. Furthermore, such sample inferences will  be important in understanding the capability of X-ray binaries to act as cosmic ray accelerators, as suggested by the recent identification of gamma-ray haloes around several Galactic X-ray binaries \citep{LHAASO2024,2024Natur.634..557A}. In turn, as the sample of jet–ISM relics grows, these structures become an in situ probe of Galactic gas, which will yield constrains on the pre- and post-shock densities, temperatures and ambient pressures, as well as the on the local sound speed and gas phase. 
Thanks to its capabilities, the SKA will move the field from serendipitous discoveries to a predictive framework for jet-driven ISM restructuring by XRBs, closing the loop between accretion physics, kinetic feedback, and Galactic ecology.

\section{Mass scaling and comparative studies}

It is generally accepted that the accretion–ejection process is scale-invariant across a wide range of compact object masses in XRBs and AGN. Nevertheless, many aspects of the underlying physics remain unresolved despite steady progress over recent decades. Current studies rely predominantly on three parameters: the estimated mass of the compact object and the (quasi-)simultaneously measured radio and X-ray luminosities in specific bands. Both observational uncertainties and sample selection criteria significantly affect the assessment of the jet–disk coupling, as expressed through the scaling relation with compact object mass. Consequently, the mathematical formulation of this relation is presently more effective as an inference framework than as a parameter estimator.

\subsection{Radio/X-ray correlations in XRBs}

Observations of black hole X-ray binaries (BHXRBs) have established a well-known correlation between the radio luminosity at 5 GHz and the X-ray luminosity in the 2–10 keV band, typically expressed as $L_R \propto L_X^b$ with $b \sim 0.5-0.7$ \citep{Gallo2003}. This relation is a defining feature of the hard accretion state, whereas in softer states the jet is strongly suppressed and the associated radio emission often falls below detectability. Despite its robustness, the correlation exhibits significant scatter, reflecting the influence of observational uncertainties, limited sample sizes, and selection effects.

An increasing number of sources deviate from the canonical relation, prompting suggestions that more than one track exists in the radio/X-ray plane. A conservative interpretation proposes two distinct branches with slopes of largely $\sim0.6$ and $\sim1.0$, respectively \citep{Gallo2012}. The shallower branch is often associated with radiatively inefficient accretion flows (RIAFs), such as advection-dominated accretion flows \citep[ADAFs;][]{Narayan1995}, in which X-ray emission scales non-linearly with accretion rate. The steeper branch may correspond to radiatively efficient regimes, potentially involving geometrically thin, optically thick accretion disks \citep{Shakura1973}, enhanced jet power \citep{Heinz2003}, or jet-dominated states \citep{Markoff2003}. Magnetically arrested disks \citep[MADs;][]{Narayan2003,Tchekhovskoy2011} have also been proposed as a mechanism for modulating jet efficiency. Some BHXRBs are observed to transition between these tracks as they evolve through different accretion states \citep{Coriat2011}, suggesting that the inflow–outflow connection depends sensitively on the accretion mode, magnetic flux, and possibly black hole spin.

Neutron star X-ray binaries (NSXRBs) exhibit intrinsically fainter radio emission, making their detection considerably more challenging. Early studies based on a limited sample suggested that NSXRBs follow a radio/X-ray correlation similar to that of BHXRBs, but with a steeper slope of $b \sim 1.4$ \citep{Migliari2006}. This steeper relation was largely interpreted as a consequence of radiatively efficient accretion, since neutron stars possess a solid surface and a boundary layer where additional energy is released. In contrast, BHXRBs are often characterized by radiatively inefficient accretion flows, such as ADAFs, in which a significant fraction of the accretion energy is advected through the event horizon rather than radiated away \citep{Narayan1995}. This fundamental distinction naturally leads to a stronger scaling of X-ray luminosity with accretion rate in NSXRBs, producing a steeper radio/X-ray correlation.

However, as larger samples have become available, an increasing number of outliers have been observed that deviate from the correlation. Recent analyses \citep{Gallo2018} indicate that, with the exception of extreme subclasses, the radio/X-ray correlations for BHXRBs and NSXRBs are statistically consistent within current uncertainties, although BHXRBs remain significantly more radio loud than their neutron star counterparts. These results suggest that the simple difference in radiative efficiency between the two classes cannot fully account for the observed behavior. Additional factors, such as neutron star magnetic field strength, spin, and boundary-layer dynamics, likely play important roles in modulating jet efficiency and shaping the observed phenomenology.

Overall, while both BHXRBs and NSXRBs exhibit radio/X-ray correlations indicative of a coupling between inflow and outflow, the differences in normalization, and scatter reflect the underlying physics of accretion and jet formation. A unified theoretical framework capable of simultaneously explaining the correlations in both classes of systems remains an open challenge, representing a key avenue for future observational and modeling efforts.

In the case of XRBs, the mass of the compact object does not explicitly enter into considerations of disk–jet coupling, as the relatively narrow mass range (typically a few solar masses) makes any dependence difficult to discern. In contrast, when comparing XRBs to AGNs, where central black hole masses span several orders of magnitude, mass plays an important role in scaling relations, influencing both accretion physics and jet properties. Consequently, while disk–jet correlations in XRBs are primarily studied in terms of luminosities and spectral states, broader frameworks, such as the fundamental plane of black hole activity (see Section~\ref{sec:FP}), are necessary to explore mass-dependent trends across the full range of compact objects.

\subsection{Fundamental plane of black hole activity}
\label{sec:FP}

The so-called "fundamental plane of black hole activity" represents an observed correlation between the black hole mass and the radio and X-ray luminosities of BHXRBs and AGNs. In its original form \citep{Merloni2003} it reads: $\log L_{\rm R} = 0.60_{-0.11}^{+0.11} \log L_{\rm X} + 0.78_{-0.09}^{+0.11} \log M_{\rm BH} + 7.33_{-4.07}^{+4.05}$, where $L_{\rm R}$ is the 5 GHz radio luminosity in units of erg\,s$^{-1}$, $L_{\rm X}$ is the 2-–10 keV band X-ray luminosity in units of erg\,s$^{-1}$, and $M_{\rm BH}$ is the black hole mass in units of $M_{\odot}$. 

Just as in the case of XRBs alone, the fundamental plane relation exhibits significant scatter, primarily due to observational uncertainties, limited sample sizes, and selection effects. Moreover, it is only valid for radiatively inefficient sources, since BHXRBs in the soft state and luminous AGNs with bolometric luminosities exceeding roughly 1\% of the Eddington limit tend to follow a slightly different scaling relation \citep{Dong2014,Gultekin2019}.

The SKA will inform the radio component of such studies. In its early-phase AA* configuration, the SKA will already surpass the sensitivity of the Karl G. Jansky Very Large Array (VLA) at relevant frequencies. For instance, 10 minutes of integration with SKA-Mid in Band 5a will achieve a Briggs-weighted sensitivity of approximately 3.4 $\mu$Jy\perbeam\ per 1 GHz-wide sub-band. This improved performance will increase the available sample size by enabling observations of fainter sources and lower-luminosity states in XRBs. Additionally, sources that are currently inaccessible to northern hemisphere radio facilities will become routine SKA targets.

The higher angular resolution of SKA (approximately 140 × 120 mas in Band 5a) will also reduce measurement uncertainties in cases where extended radio emission contaminates flux density determinations, thereby improving the accuracy of radio luminosity estimates. Even in its initial \AAstar\ configuration, SKA will enable a more systematic investigation of mass-scaling relations by providing self-consistent, high-precision measurements of radio flux densities. This will minimize the need for band conversions or cross-calibration between different radio arrays, allowing for a more robust characterization of the fundamental plane across a broad range of compact sources.

Beyond improving measurement precision, SKA’s enhanced sensitivity and survey capabilities will allow the detection of statistically significant samples of low-luminosity AGNs and quiescent XRBs, regimes that are currently poorly sampled. This will help disentangle intrinsic scatter from observational biases, offering the opportunity to test whether the fundamental plane represents a single, universal scaling relation or a set of distinct correlations corresponding to different accretion regimes. SKA’s contribution will be important in unifying the study of black hole activity across the entire mass spectrum, from stellar-mass BHXRBs to supermassive AGNs, providing a clearer empirical foundation for theories of accretion and jet production.

\section{Jets at the highest accretion rates}

Ultraluminous X-ray sources (ULXs) are non-nuclear, extragalactic compact objects with X-ray luminosities (\(L_{\mathrm{X}} \gtrsim 10^{39}\)~erg~s\(^{-1}\)), i.e., exceeding the Eddington limit for a 10~\(M_{\odot}\) black hole \citep{Kaaret2017, King2023}. Initially interpreted as hosts of intermediate-mass black holes (IMBHs; \(10^{2} - 10^{5}\,M_{\odot}\)) accreting below the Eddington rate \citep{Colbert1999, Miller2004}, particularly for hyperluminous cases (\(L_{\mathrm{X}} > 10^{41}\)~erg~s\(^{-1}\); \citealt{Webb2010, MacKenzie2023}), most ULXs are now understood to be stellar-mass systems accreting at super-Eddington rates, facilitated by optically thick outflows or geometric beaming \citep{Fabrika2015}. This interpretation is supported by dynamical mass estimates, detections of ULXs hosting stellar-mass BHs \citep{Motch2014, Shen2015}, and the discovery of multiple pulsating ULXs confirming NS accretors (e.g. \citealt{Bachetti2014, Israel2017a, Israel2017b, Carpano2018, Sathyaprakash2019}). ULXs therefore provide laboratories to study accretion at the highest rates.

Recent multi-wavelength studies revealed signatures of collimated outflows, for example, the 300-parsec jet-inflated bubble in NGC 7793, reported by \citet{Pakull2010}. Using optical imaging and spectroscopy (shock velocities $\sim$~250 \rm{km\,s}$^{-1}$) and X-ray hotspots, they inferred a jet power of the order of a few ${\times}10^{40}$ $\rm{erg s^{-1}}$.~However, the absence of high-sensitivity, high-resolution radio data leaves key aspects unresolved. Without radio observations, it is difficult to map the synchrotron emitting regions, to measure spectral indices, or to trace magnetic field structure.~With its unprecedented sensitivity and resolution, SKA will be able to image the nonthermal radio lobes (and similar bubbles) in exquisite detail, mapping low surface-brightness synchrotron emission, measuring spectral breaks and polarized emission \citep{SKA_polarized_ULXs}, and thus opening a window into the physics of ULX jets that is currently only partially accessible. For this purpose, SKA's sensitivity to extended and low-surface brightness emission, which is now routinely highlighted by its pathfinder telescopes, will be essential.

High-resolution radio imaging with facilities such as the VLA and EVN has revealed compact and, in several cases, multi-component radio structures within ULX-powered nebulae. These observations are consistent with the presence of collimated jets and recurrent ejection events. Holmberg~II~X-1 provides a prototypical example: VLBI/EVN and VLA observations uncovered a triple radio morphology and evolving compact components interpreted as jet knots or recurrent outflows \citep{Cseh2014}. Similar evidence for parsec-scale jet features and steady radio cores has been reported in other ULXs, including sources exhibiting mas-scale inner jets \citep[e.g.][]{Yang2023}. These findings indicate that collimated, possibly relativistic jets can coexist with super-Eddington accretion flows, providing crucial insight into jet formation and feedback in extreme accretion regimes.~Some ULXs have been observed to exhibit radio variability and steep spectral indices (e.g. NGC 5408 X-1,~Holmberg II X-1) consistent with non-thermal jet or lobe emission, but there is as yet no secure detection of fast transient radio ejections analogous to those seen in Galactic BH XRBs.

The discovery of pulsations in several ULXs (demonstrating neutron-star accretors) raised the question of whether such systems may launch detectable radio jets. These pulsating ULXs hosts neutron stars with properties akin to those in Galactic high-mass X-ray binaries: young, relatively slowly spinning (periods $>1$ second), and strongly magnetized ($B>10^{12}$ G). Studies of these Galactic counterparts, across a wide range of accretion rates (including $>10^{39}$ erg\,s$^{-1}$), reveal that the jets launched by such strongly-magnetized and slowly spinning NSs are orders of magnitude fainter than their BH siblings \citep[e.g.,][]{2022MNRAS.516.4844V}. A 2024 study concluded that current facilities find little radio emission from many NS-ULXs \citep{Panurach2024}. Even if the observed correlation between X-ray and radio luminosity for strongly-magnetized NSs extends into the extremely super-Eddington regime ($\sim 10^{41}$ erg\,s$^{-1}$), sensitivities substantially below $1$ $\mu$Jy would be needed to detect their radio luminosity at Mpc distances. As confusion and host galaxy emission may further complicate radio detection, it appears unlikely that many pulsating ULXs will be detectable with SKA. 

The advent of the SKA will mark a transformative step in this field. With its unprecedented sensitivity (sub-$\mu$Jy levels), wide frequency coverage, and high angular resolution, SKA will enable systematic surveys of ULX radio counterparts across diverse host environments. It will detect faint, steady radio cores from currently undetected systems, trace variability associated with recurrent jet ejections, and resolve the morphology and energetics of ULX bubbles on sub-parsec scales. Furthermore, multi-epoch SKA monitoring will constrain jet duty cycles and kinetic feedback efficiencies, offering new tests for accretion–ejection coupling in the super-Eddington regime. In synergy with contemporaneous X-ray facilities (e.g. \textit{NewAthena}, and \textit{XRISM}), SKA observations will allow a direct comparison of accretion states and jet power, providing a complete picture of ULX feedback and its role in galaxy evolution. Furthermore, in combination with the Extremely Large Telescope, the accretion and ejection properties of ULXs can be combined with a systematic exploration of the binary and donor star properties of these enigmatic systems. 

\section{The black hole population of the Milky Way}
Population synthesis models predict $\sim$ $10^{7}-10^{8}$ stellar-mass black holes in the Milky Way \citep[e.g.,][]{Agol2002,Wiktorowicz2019}.~Only about 20 dynamically confirmed BH XRBs and $\sim$ 100 candidates are known today -- a tiny fraction of the predicted total.~Nearly all known BHs are in transient systems, bright in X-rays only during rare outbursts; for $>$ 99{\%} of their lifetimes, they are in quiescence. Quiescent stellar-mass black holes are expected to produce faint, compact radio jets even at extremely low accretion rates. Observations of nearby quiescent XRBs such as A0620–00, V404~Cyg, and XTE~J1118+480 have revealed persistent radio emission at the $\mu$Jy level, consistent with synchrotron radiation from weak, steady outflows \citep{Gallo2005, Gallo2006,Gallo2014}. Using the empirical radio--X-ray correlation established for BH XRBs \citep{Gallo2003, Plotkin2021}, extrapolated to quiescent luminosities ($L_{\rm X} \sim 10^{30-33}$~erg~s$^{-1}$), the expected radio luminosities are $L_{\nu} \sim 10^{26-28}$~erg~s$^{-1}$~Hz$^{-1}$, corresponding to flux densities of a few to tens of~$\mu$Jy for sources located within 1--5~kpc. These values are consistent with direct detections obtained using the VLA and other sensitive interferometers, confirming that even in quiescence, black holes remain weakly radio-loud.~The unprecedented sensitivity and survey speed of the Square Kilometre Array (SKA) will revolutionize the search for faint accreting black holes in our Galaxy. The SKA-Mid all-sky continuum surveys, with rms sensitivities of a few~$\mu$Jy, are expected to detect hundreds to thousands of the nearest quiescent black hole binaries -- although confirming their nature will remain a challenge without extensive multiwavelength data. Deeper targeted observations and Galactic plane surveys with the full SKA, achieving sensitivities of $\sim0.1~\mu$Jy~rms, could probe to several kiloparsecs and potentially reveal $10^{4}$--$10^{5}$ faint black hole systems, depending on assumptions about jet beaming, binary duty cycles, and spatial distribution \citep[e.g.][]{Maccarone2015, Fender2017}.

\section{Jets as probes}

The radio emission from X-ray binary jets can also be used to determine their astrometric properties; their proper motions \citep[e.g][]{Mirabel2001}, distances \citep[e.g][]{Bradshaw1999}, and in some cases even orbital motion \citep[e.g][]{Miller-Jones2021}. The proper motions and distances can be used in conjunction with optical measurements of the systemic radial velocity of the donor star to determine the full three-dimensional motion of the system, and hence its Galactocentric orbit, providing some of the few direct observational constraints on how black holes form \citep{Atri2019,Nagarajan2025}. Astrometric measurements of the jet motion during the binary orbit, in conjunction with Gaia astrometry or radial velocity curves for the donor star can uniquely solve for the orbital parameters and the component masses.

\begin{figure}[h]
    \centering
	\includegraphics[width=0.8\columnwidth]{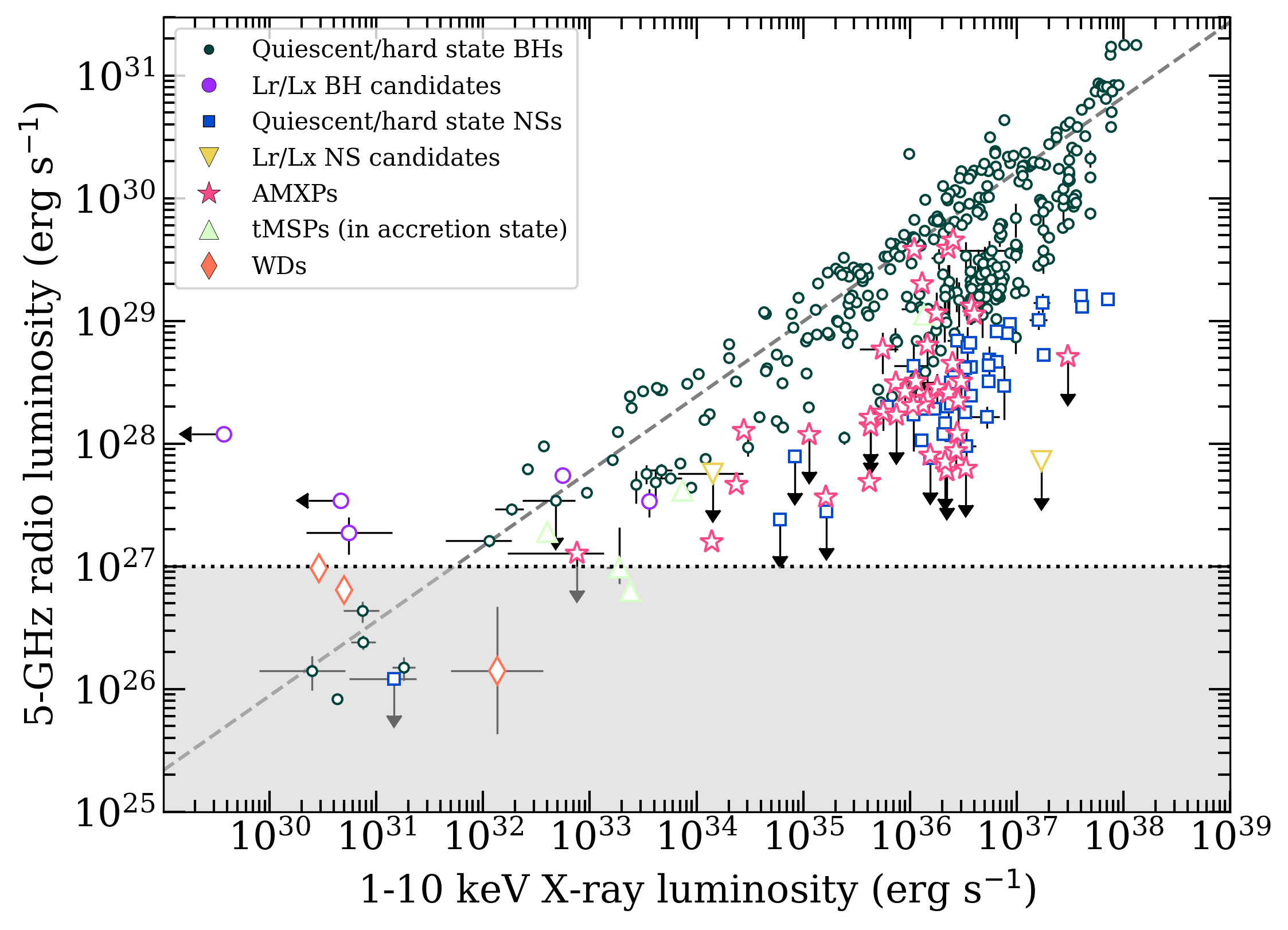}
    \caption{Radio/X-ray luminosity plane, showing the measured positions of a sample of black hole and neutron star X-ray binaries, indicating the (unshaded) region accessible to SKA-VLBI astrometric observations (detection threshold 10\,$\mu$Jy) for a source at 4\,kpc \citep{Bahramian2022}.}
    \label{fig:astrometry_lr_lx}
\end{figure}

The limiting factor for current astrometric campaigns has been sensitivity; BH X-ray binaries spend most of their time in a faint, quiescent state, rendering them inaccessible to astrometric VLBI observations. The microJy-level sensitivity of VLBI arrays that include the phased-up SKA-Mid core, coupled with astrometric precision at the few microarcsecond level \citep{Li2024} will open up a wider range of systems to astrometric study (Fig.~\ref{fig:astrometry_lr_lx}), as discussed by \citet{SKA_XRBs_VLBI}. While the majority of quiescent systems will remain too faint for radio detection, the increased sensitivity will enable astrometric measurements for a longer period of the outburst rise and decay, extending the time baseline for astrometric campaigns. Furthermore, the increased sensitivity will allow for the use of fainter background calibrator sources, reducing the systematic uncertainties associated with the phase referencing calibrator throw.

Astrometric campaigns with phased SKA-Mid as a VLBI element should begin as soon as radio emission is detected in the rising phase of the outburst, and continue until the outburst is over and the source has faded below the detection threshold. Observations should be timed to coincide with the times of maximum parallax displacement in RA and Dec co-ordinates, with 5 optimally-timed measurements over 1 year (if possible) being sufficient to measure both the parallax and proper motion of newly-detected sources. Microarcsecond astrometry would permit the measurement of orbital displacements for longer-period systems (of order days), especially high-mass X-ray binaries, in which the semi-major axis of the jet-producing compact object was a significant fraction of the orbital separation. A more quantitative discussion of the feasibility of such measurements can be found in \citet{Lin01.2026.SKA}.

\section{Comparing Black Holes and Neutron Stars: Dissecting the Physics of Jet Launching}

The SKA will revolutionize comparative jet studies of black holes and neutron stars by providing unprecedented sensitivity.~SKA will detect radio jets from faint neutron star systems that are currently below detection limits. This will help establish whether all accreting NSs launch jets and how jet power scales across different magnetic fields and spin rates. Its rapid survey and monitoring capabilities will allow tracking of state transitions (hard → soft) in X-ray binaries in real time, revealing how jets turn on/off and evolve in response to changing accretion conditions.~With wide-band and polarimetric sensitivity, SKA will map magnetic-field geometry and particle acceleration zones in jets, constraining how magnetic flux and spin interplay to drive outflows.~Joint SKA + X-ray observatory campaigns (e.g. with NewAthena, XRISM) will directly correlate disc variability with jet response, tightening constraints on jet-launching physics.

The comparison between jets of BH and NS X-ray binaries often focuses on the presence and effects of the neutron star's spinning stellar surface and magnetic field on jet formation. At its most basic level, these prevent the operation of certain jet launch mechanisms \citep[e.g., the Blandford-Znajek mechanism), while enabling others (for instance, powered by magnetic fields lines of the NS itself;][]{Parfrey2024,Das2024}. To understand this difference between NS and BH jets (and jets of different types of NSs), the SKA will enable the systematic application of various techniques currently used in only a few sources. With its sensitivity, for instance, the study of the faint jets of strongly-magnetized and slowly-spinning NS can be extended to larger distances and lower accretion rates; because roughly half of all X-ray binaries host such a NS, the SKA will provide a greatly enhanced sample of sources where the jet properties can be studied as a function of NS properties \citep{2022MNRAS.516.4844V}. The wide frequency band and excellent sensitivity of the SKA will also enable more systematic timing studies of X-ray binary jets across NS and BH systems, tracing how variability propagates down the jet as a constraint on jet geometry and speed \citep{2021MNRAS.504.3862T, 2024Natur.627..763R}.

Moving even lower on the mass scale, the sensitivity of SKA will also allow systematic exploration of the faint radio emission known to exist in accreting white dwarf systems \citep[e.g][]{Koerding2008, Coppejans2015, Coppejans2016}, determining whether these accreting systems launch jets, and if so, how their properties scale with mass. The high space density of cataclysmic variables means that SKA surveys will pick up many such systems, allowing statistical samples that are not achievable for their rarer, more massive counterparts in the NS and BH XRBs.

\section{Summary}

The primary benefit that the SKA will bring to the field of accreting compact objects is its significantly enhanced sensitvity relative to the current generation of interferometers. This enables the extension of the rich studies of accreting Galactic black hole systems to their fainter neutron star (and even white dwarf) counterparts, as well as to rare, high accretion rate systems in external galaxies. Jet studies in these poorly-constrained regions of parameter space will advance our understanding of the mechanisms required to launch and accelerate jets, determine how they are linked to the underlying accretion flow, and quantify their feedback effect on the surroundings. SKA surveys may uncover previously undetected quiescent black holes, refining our understanding of the Galactic black hole population, and hence of stellar and binary evolution. Finally, high-precision astrometry with the phased SKA as a component of global VLBI arrays will probe black hole formation.

Since accreting stellar-mass compact objects vary in luminosity by several orders of magnitude over the course of an outburst, the sensitivity of SKA will not be required for the full duration of a monitoring campaign. We encourage the community to ensure that SKA is complemented by existing lower-sensitivity facilities that have the capacity to conduct regular, intensive monitoring of important targets, in the same way that 1m, 2m, and 4m-class telescopes remain important for transient science at optical wavelengths. While such facilities may be under funding pressure, they will continue to play a critical role in providing high-cadence monitoring of the brighter phases of XRB duty cycles \citep{Fender2023}.

\bibliographystyle{abbrvnat-maxbibnames4}
\bibliography{chapter}

\end{document}

%% file: journal-names.tex
\newcommand{\actaa}{Acta Astron.} 
\newcommand{\araa}{ARA\&A} 
\newcommand{\aar}{A\&ARv} 
\newcommand{\aapr}{A\&ARv} 
\newcommand{\ab}{Astrobiol.} 
\newcommand{\aj}{AJ} 
\newcommand{\apj}{ApJ} 
\newcommand{\apjl}{ApJL} 
\newcommand{\apjs}{ApJSS} 
\newcommand{\ao}{Appl. Opt.} 
\newcommand{\apss}{Astro. \& Space Sci.} 
\newcommand{\aap}{A\&A} 
\newcommand{\aaps}{A\&AS.} 
\newcommand{\baas}{Bull. Am. Astron. Soc.} 
\newcommand{\caa}{Chinese A\&A} 
\newcommand{\cjaa}{Chinese J. A\&A} 
\newcommand{\cqg}{Class. Quantum Gravity} 
\newcommand{\gal}{Galaxies} 
\newcommand{\gca}{Geo. Cosmo. Acta} 
\newcommand{\icarus}{Icarus} 
\newcommand{\jcap}{JCAP} 
\newcommand{\jgr}{J. Geophys. Res.} 
\newcommand{\jgrp}{J. Geophys. Res. Planets} 
\newcommand{\jqsrt}{J. Quant. Spectrosc. Radiat. Transf.} 
\newcommand{\memsai}{Mem. SAIt} 
\newcommand{\mnras}{MNRAS} 
\newcommand{\nat}{Nature} 
\newcommand{\nastro}{Nat. Astron.} 
\newcommand{\ncomms}{Nat. Commun.} 
\newcommand{\nphys}{Nat. Phys.} 
\newcommand{\na}{New Astron.} 
\newcommand{\nar}{New Astron. Rev.} 
\newcommand{\physrep}{Phys. Rep.} 
\newcommand{\pra}{Phys. Rev. A} 
\newcommand{\prb}{Phys. Rev. B} 
\newcommand{\prc}{Phys. Rev. C} 
\newcommand{\prd}{Phys. Rev. D} 
\newcommand{\pre}{Phys. Rev. E} 
\newcommand{\prx}{Phys. Rev. X} 
\newcommand{\prl}{Phys. Rev. Let.} 
\newcommand{\psj}{Planet. Sci. J.} 
\newcommand{\planss}{Planet. Space Sci.} 
\newcommand{\pnas}{Proc. Natl Acad. Sci. USA} 
\newcommand{\procspie}{Proc. SPIE} 
\newcommand{\pasa}{PASA} 
\newcommand{\pasj}{PASJ} 
\newcommand{\pasp}{PASP} 
\newcommand{\rmxaa}{RMXAA} 
\newcommand{\sci}{Science} 
\newcommand{\sciadv}{Sci. Adv.} 
\newcommand{\solphys}{Sol. Phys.} 
\newcommand{\sovast}{Soviet Ast.} 
\newcommand{\ssr}{Space Sci. Rev.} 
\newcommand{\uni}{Universe} 